\documentclass[12pt,preprint]{aastex}
\usepackage{natbib,graphics,graphicx}

\slugcomment{3 April 2009 version; Accepted by AJ}

\shorttitle{Astrometric Redshifts for Quasars}
\shortauthors{Kaczmarczik et al.}

\begin{document}

\title{Astrometric Redshifts for Quasars}

\author{Michael C. Kaczmarczik, Gordon T. Richards, Sajjan S. Mehta}
\affil{Department of Physics, Drexel University, Philadelphia, PA 19104}

\author{David J. Schlegel}
\affil{Lawrence Berkeley National Lab, 1 Cyclotron Road, MS 50R5032, Berkeley, CA 94720}

\begin{abstract} 

The wavelength dependence of atmospheric refraction causes
differential chromatic refraction (DCR), whereby objects imaged at
different optical/UV wavelengths are observed at slightly different
positions in the plane of the detector.  Strong spectral features
induce changes in the effective wavelengths of broad-band filters that
are capable of producing significant positional offsets with respect
to standard DCR corrections.  We examine such offsets for
broad-emission-line (type 1) quasars from the Sloan Digital Sky Survey
(SDSS) spanning $0<z<5$ and an airmass range of 1.0 to 1.8.  These
offsets are in good agreement with those predicted by convolving a
composite quasar spectrum with the SDSS bandpasses as a function of
redshift and airmass.  This astrometric information can be used to
break degeneracies in photometric redshifts of quasars (or other
emission-line sources) and, for extreme cases, may be suitable for
determining ``astrometric redshifts''.  On the SDSS's southern
equatorial stripe, where it is possible to average many multi-epoch
measurements, more than 60\% of quasars have emission-line-induced
astrometric offsets larger than the SDSS's relative astrometric errors
of 25-35 mas.  Folding these astrometric offsets into photometric
redshift estimates yields an improvement of 9\% within $\Delta
z\pm0.1$.  Future multi-epoch synoptic surveys such as LSST and
Pan-STARRS could benefit from intentionally making $\sim10$
observations at relatively high airmass (${\rm AM}\sim1.4$) in order
to improve their photometric redshifts for quasars.



\end{abstract}


\section{Introduction}

The next generation of large-area survey facilities (e.g., Pan-STARRS,
LSST, DES, VISTA/VST; \citealt{kab+02,lsst08,des05,anp+07}) will image
orders of magnitude more objects than it will be possible to obtain
spectra for.  As a result, robust determination of redshifts from
photometric data is of crucial importance to these projects.  While
the distinctive ``4000\AA$\;$break'' facilitates photometric redshifts
(``photo-$z$'s'') for galaxies \citep[e.g.,][]{ccs+95,olc+08}, the
majority of quasars ($z\lesssim2.2 $) lack such a strong spectral
feature at observed-frame optical wavelengths, making the
determination of photometric redshifts for low-$z$ quasars more
challenging.  While it has been proven to be possible to determine
photo-$z$'s of quasars from the changes that their emission lines
induce in their broad-band photometry
\citep[e.g.,][]{wmr01,rws+01,bcs+01,wrs+04,bbm+08} and from spectral
energy distribution (SED) fitting \citep[e.g.,][]{bba+06,rbo+08}, the
accuracy of these photo-z's does not match that of galaxies and often
suffers from catastrophic failures ($|\Delta z|>0.3$).

Here we introduce a powerful new tool to aid redshift determination
for emission-line objects (such as quasars) by taking advantage of the
spectroscopic properties of the Earth's atmosphere.  Light rays from
extraterrestrial sources are bent according to Snell's law as they enter the Earth's atmosphere from the vacuum of space.  As a result,
except at the zenith, a celestial source observed from the Earth will
appear higher in the sky than it actually is.  The magnitude of this
deflection depends on the index of refraction in air and the photon's
angle of incidence.  Since the index of refraction of air is a
function of wavelength, light rays passing through the atmosphere
undergo dispersion, whereby shorter wavelength photons are bent more
than longer wavelength photons.  \citet{fil82} provides an in-depth
discussion of this effect, reminding observers of the rather large
magnitude of this effect even at moderate airmasses and the need to
perform spectroscopy at the ``parallactic angle''.


This effect, known as differential chromatic refraction (DCR; e.g,
\citealt{pmh+03}), is normally a source of nuisance for multi-band
astrometry.  Indeed oftentimes additional lenses are used to form an
atmospheric dispersion corrector (ADC) to compensate for DCR in
hardware, such as for the 2dF system \citep{2df} on the
Anglo-Australian Telescope.  However, rather than being a nuisance,
DCR can be used as a unique tool for estimating redshifts of
emission-line objects.  Software-based astrometric corrections for DCR
are generally computed as a function of broad-band flux ratio (i.e.,
color), whereas the actual DCR in emission-line objects instead
depends on the distribution of flux {\em within} the bandpass.  The
difference between the expected and observed astrometric displacements
due to DCR enables determination of {\em astrometric redshifts} (or
astro-$z$'s) for strong emission-line objects.  This discrepancy can
be used separately from, or in conjunction with, more traditional
photo-$z$ estimates to provide redshifts from imaging data.  Indeed,
contrary to conventional wisdom, the next generation of multi-epoch
imaging surveys may want to perform a fraction of their observations
at moderately high airmass in order to take full advantage of this
effect.



The sections of this paper are as follows.  Section~2 discusses DCR
from a theoretical perspective, while \S~3 considers DCR observations
from both single- and multi-epoch quasar data from the Sloan Digital
Sky Survey (SDSS; \citealt{yaa+00}).  Section~4 examines how these
measurements can be used to improve photometric redshift estimation
for quasars.  Finally, \S~5 discusses some avenues for further work
and \S~6 presents our conclusions.

\section{Differential Chromatic Refraction}

\subsection{Theory}
\label{sec:theory}

\begin{figure}
\plottwo{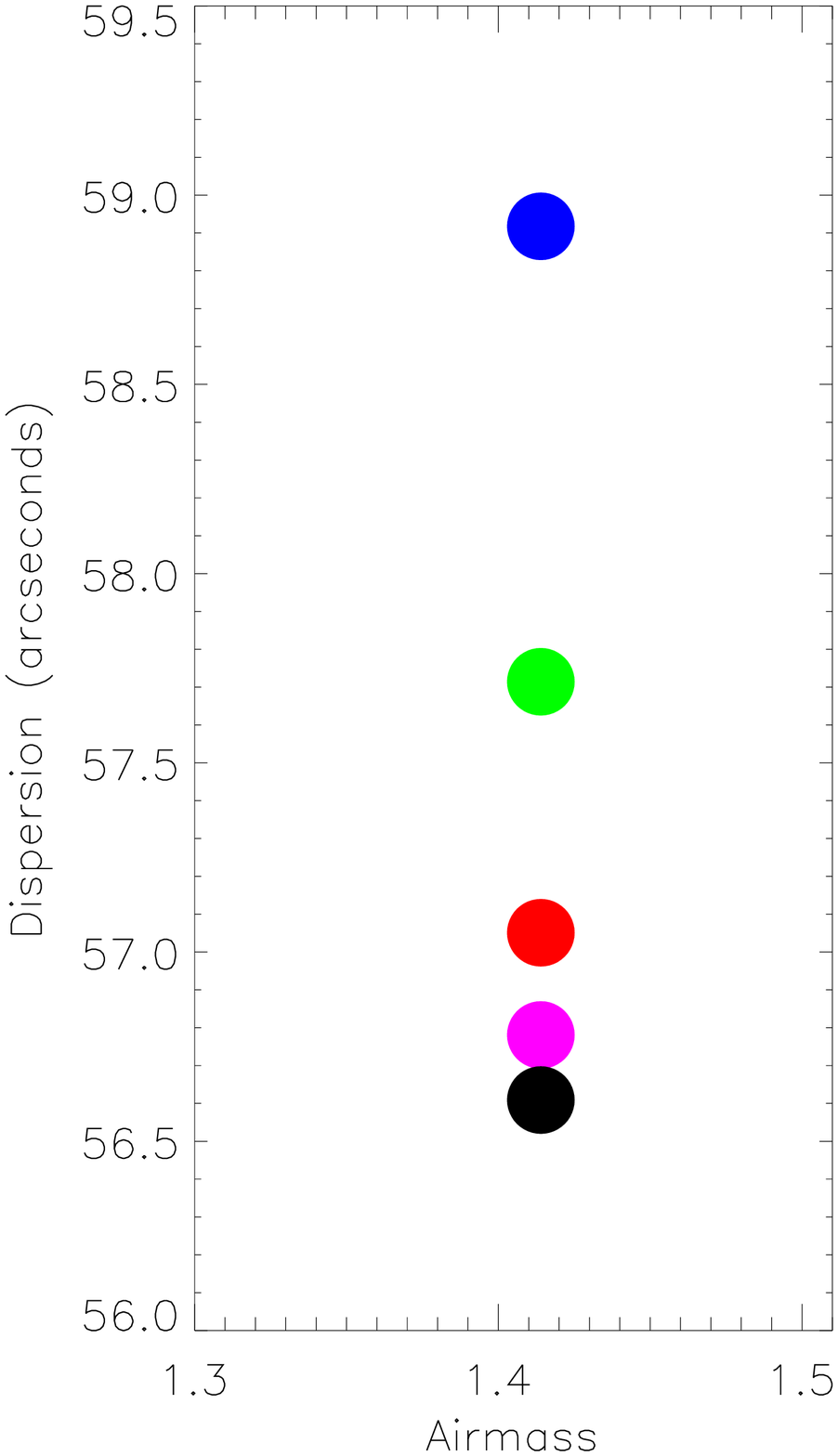}{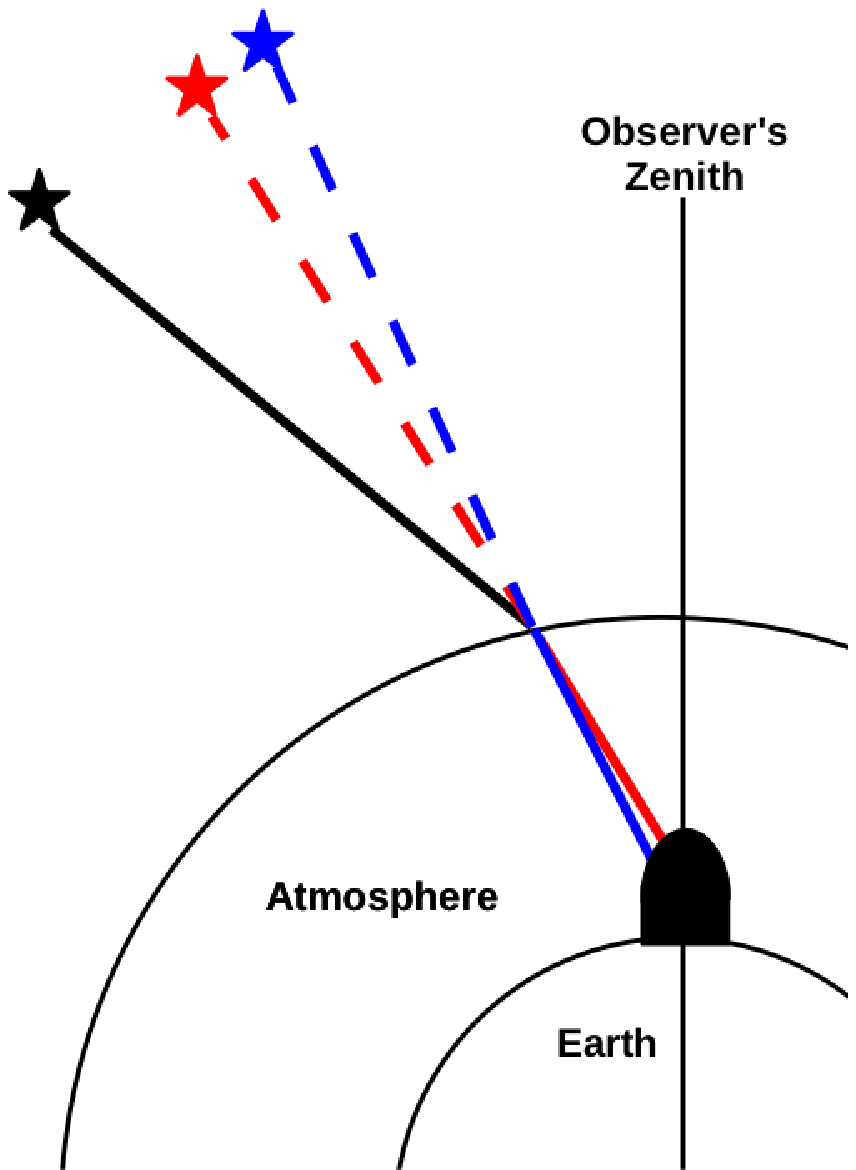}
\caption{{\em Left:} Differential chromatic refraction (DCR) for a
  flat-spectrum object observed in the SDSS photometric system at a
  zenith angle of 45 degrees (${\rm AM} = 1.414$).  The color coding
  is $u=$blue, $g=$green, $r=$red, $i=$magenta, $z=$black.  Objects
  appear higher in the sky when observed in blue bandpasses than in
  red bandpasses.  {\em Right:} DCR schematic example.  The solid
  black line indicates the incoming multi-chromatic light rays.  The
  solid red and blue lines indicate the differential chromatic
  refraction of the incoming beam, with blue light rays being bent
  more than red.  The dashed blue and red lines indicate the apparent
  location on the sky of the object as seen by the blue and red
  filters.
\label{fig:dcr}}
\end{figure}

Differential chromatic refraction is a fact of life for any multi-band
astronomical imaging survey.  Because of the wavelength dependence of
the refractive index of air, a source observed with a blue filter will
appear slightly higher in the sky than the same source observed
through a red filter.  Figure~\ref{fig:dcr} illustrates this effect.
The left-hand panel indicates the magnitude of the effect as a
function of wavelength; here for the SDSS filter set.  The right-hand
panel of Figure~\ref{fig:dcr} illustrates why the blue image is higher
in the sky than the red image.  It is clear that refraction is a
potentially large source of astrometric error and both it and DCR must
be carefully corrected (either in hardware with an ADC or in software)
in order to produce accurate astrometric solutions.


Here we explicitly consider the absolute theoretical wavelength
depedence of the deflection of photons from astronomical sources. The
angular deflection of an incoming photon by the Earth's atmosphere is
given by
\begin{equation}
R \simeq R_0\tan(Z),\label{eq:R}
\end{equation}
where $Z$ is the angle from the zenith (airmass, ${\rm AM}=\sec[Z]$;
good for $Z<80$) and $R_0$ is determined from the index of refraction,
$n$, as
\begin{equation}
R_0 = \frac{n^2-1}{2n^2}
\end{equation}
\citep[e.g.,][]{cox00}.  The wavelength dependence of the index of
refraction of air can be described by \begin{eqnarray}
  [n({\lambda})-1]10^6=64.328 + \frac{29498.1}{146-(1/\lambda)^2}
  +\frac{255.4}{41-(1/\lambda)^2},\label{equ:n_lambda} \end{eqnarray}
where $\lambda$ is expressed in microns \citep[e.g.,][]{fil82}.  Thus,
a flat-spectrum source\footnote{We use the nomenclature
  $f_{\nu}\propto{\nu}^{\alpha}$, where a flat-spectrum source has
  $\alpha_{\nu}=0$.} observed at AM=1.414 ($Z=45$) in the SDSS $r$
band (6165\AA) will appear $57\farcs06$ higher in the sky (ignoring
temperature, pressure, water vapor, and altitude effects) than it
would in the absence of Earth's atmosphere; see Figure~\ref{fig:dcr}.

To determine the DCR for any given object, it is sufficient to know
the effective wavelength of the object within a given bandpass.
Following \citet{sgh83}, effective wavelengths are given by
\begin{eqnarray}
\lambda_{eff}=\bar{\lambda_{j}}(z)= \exp{\frac{\int^\infty_0 f_{\lambda(1+z)}S_j(\lambda)ln(\lambda)\,d\lambda}{\int^\infty_0
f_{\lambda(1+z)}S_j(\lambda)\,d\lambda}}.\label{eq:lameff}
\end{eqnarray}
The subscript $j$ represents a given bandpass, $u,g,r,i$ or $z$
\citep{fig+96}. The incident flux at a given wavelength and redshift
is indicated by $f_{\lambda(1+z)}$ while $S_j(\lambda)$ is the
response of the bandpasses\footnote{See
  http://www.sdss.org/dr6/instruments/imager/.}.  The transmission
curves, $S_j(\lambda)$, for the SDSS system include atmospheric
extinction at the mean airmass of the SDSS survey, ${\rm AM}=1.3$.
For a flat-spectrum source, the nominal effective wavelengths of the
SDSS bandpasses are 3551, 4686, 6165, 7481, and 8931 \AA, respectively
for $u$, $g$, $r$, $i$, and $z$.  For a power-law continuum with
spectral index $\alpha_{\nu}=-0.5$, characteristic of quasars, the
effective wavelengths are 3541, 4653, 6147, 7461, and 8904,
respectively \citep{rfs+01}.  Using these effective wavelengths, the
DCR can be determined from Equations~1--3.

\subsection{DCR for Broad Emission-Line Quasars}

In reality, the effective wavelength will depend on the more
complicated wavelength dependence for the SEDs of real astronomical
sources.  As broad-band photometry trades a loss of information
(integration of a spectral energy distribution over $\sim1000{\rm
  \AA}$) for a gain in signal-to-noise per unit time, the information
from spectral features narrower than the photometric bandpasses is
marginalized.  Quasar emission lines, however, are sufficiently broad
and strong that their effects are quite noticeable both on their
colors (e.g., \citealt{cv90,rfs+01}), and their broad-band magnitudes
\citep{rsf+06}.  These emission lines can similarly shift the
effective wavelength of a photometric bandpass.  This shift occurs
because the effective wavelength depends not only on the filter's
transmission properties, but also on the distribution of the source's
flux within the bandpass.  Indeed, while astronomers are frequently
tempted to assume a uniform transformation between broad-band
photometric systems, those transformations are, in fact, strongly
dependent on the (observed frame) spectral energy distribution of the
objects in question \citep[e.g.,][]{fig+96,jsr+05}.

\begin{figure}
\plotone{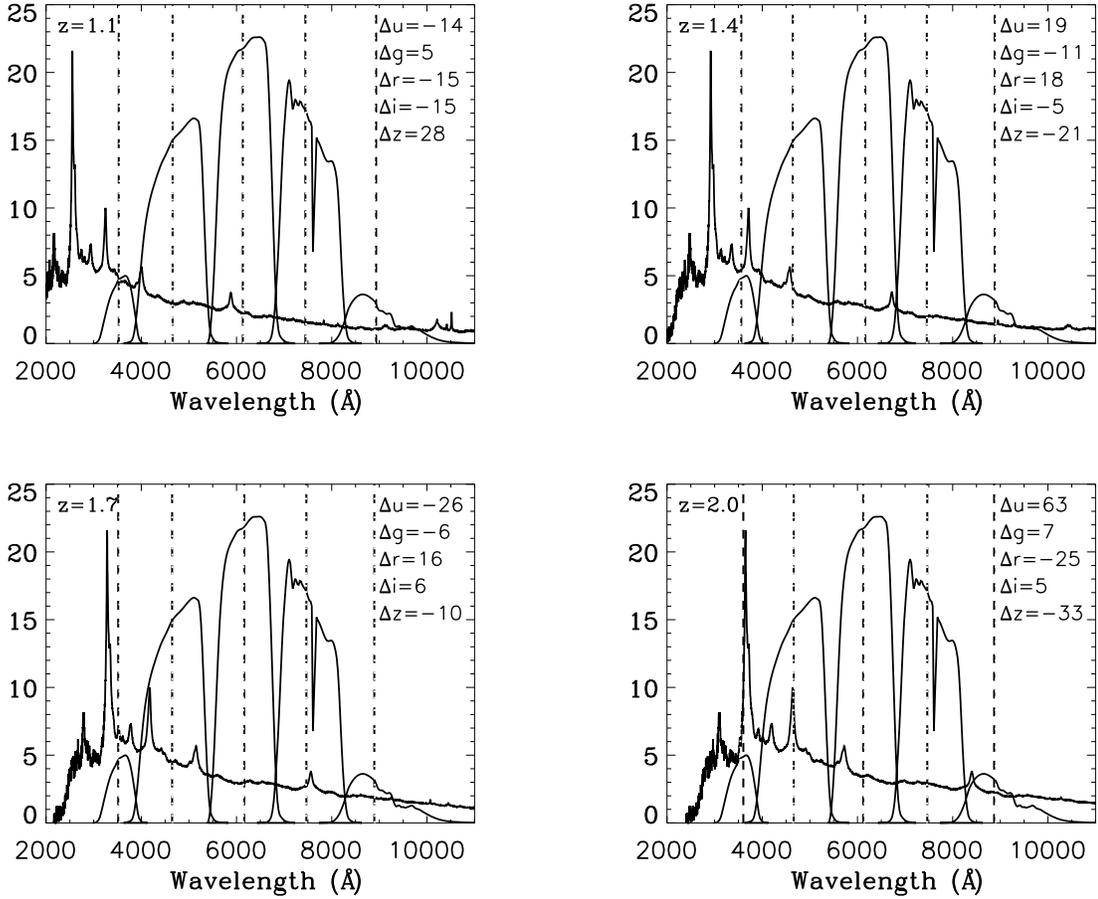}
\caption{Change in effective wavelength as a function of redshift for
  a composite quasar SED \citep{vrb+01} as compared to a power-law
  continuum source. The panels show quasars at redshifts $1.1, 1.4,
  1.7,$ and $2.0$, illustrating the effect of strong lines moving from
  the blue to the red side of a bandpass (e.g., \ion{C}{4} in $g$ for
  $z=1.1$ and $1.4$ and Lyman-$\alpha$ in $u$ for $z=1.7$ and $2.0$).
  In the upper right-hand corner of each panel, the wavelength offset
  (in Angstroms) with respect to the nominal effective wavelength is
  given for each bandpass.
\label{fig:examples}}
\end{figure}

\begin{figure}
\plotone{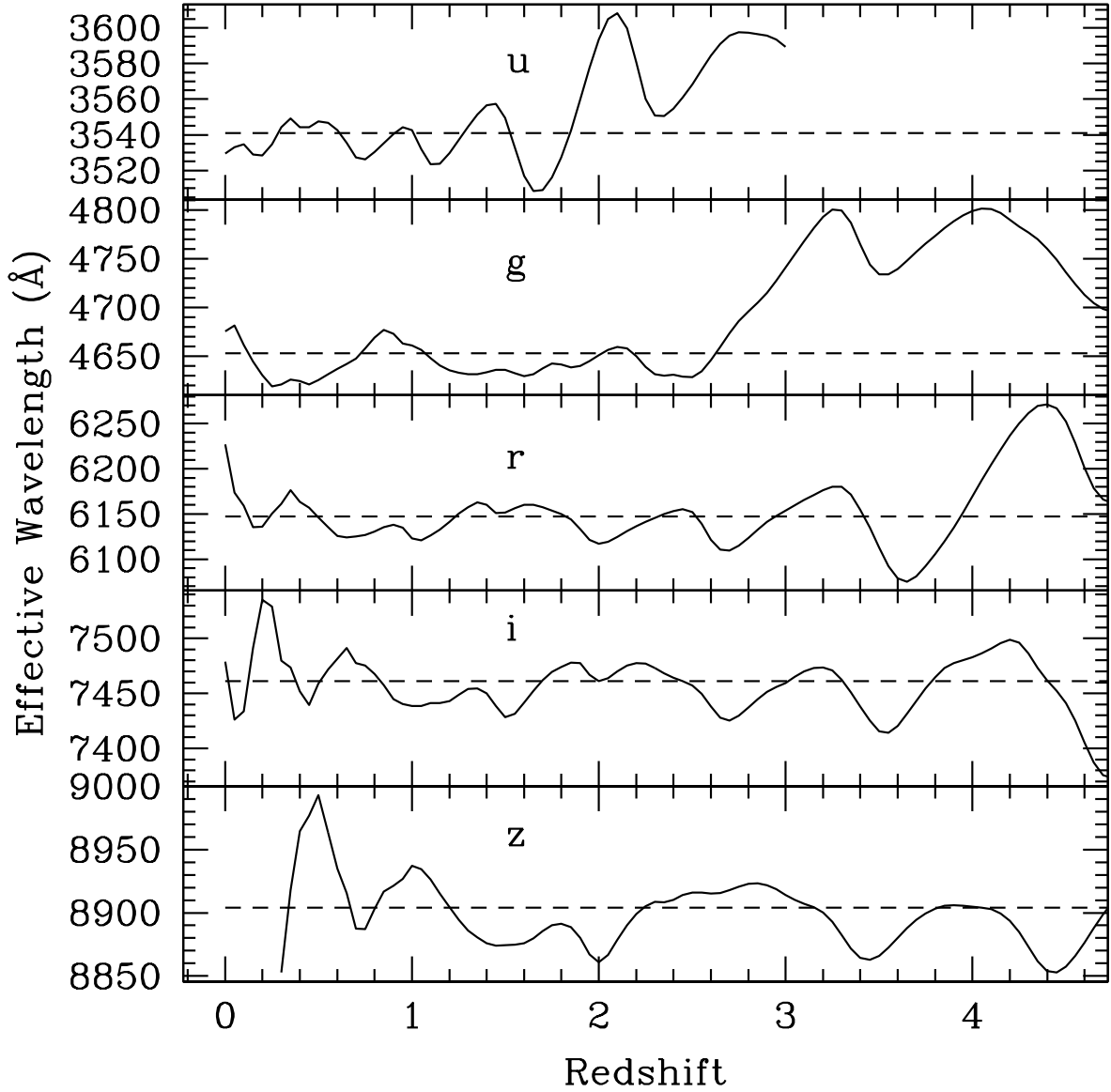}
\caption{Expected change in effective wavelength for a quasar SED as a 
function of redshift.  As a strong emission line passes through a
  given filter, the effective wavelength first shifts to the blue, then
  to the red, then returns to the standard value as the line exits the
  filter.  Each panel shows one of the $ugriz$ filters
  and its nominal expected wavelength (dashed) as compared with the
  predicted expected wavelength for a composite quasar spectrum
  (solid).  }\label{fig:lameff}
\end{figure}

In Figure~\ref{fig:examples} we illustrate the effect of quasar
emission lines on the effective wavelengths within the SDSS broad-band
photometric system.  Shown is the \citet{vrb+01} SDSS composite quasar
spectrum at four redshifts where the Ly$\alpha$ or \ion{C}{4} emission
lines lie at the red or blue edge of the $u$ filter.
Figure~\ref{fig:lameff} shows the full redshift dependence of the
effective wavelength for the mean quasar; here we have convolved the
mean quasar spectrum with the SDSS bandpasses according to
Equation~\ref{eq:lameff}.  Compared to a power-law continuum source,
the effective wavelength of the SDSS bandpasses can change by as much
as 150\,\AA\ for quasars.  (For $z\lesssim0.3$ and $z\gtrsim3.25$ the
\citealt{vrb+01} composite spectrum does not fully span the $z$ and
$u$ bands, respectively and the expected changes in effective
wavelength cannot be determined from the composite quasar spectrum.)
We can use these effective wavelengths to determine the magnitude of
the DCR for emission-line objects as a function of redshift, but first
we need to understand how DCR corrections are applied, particularly in
the case of SDSS astrometry.



\subsection{SDSS Astrometry}

As can be seen from Figure~\ref{fig:dcr}, if no DCR corrections were
applied, then potentially large positional differences between
bandpasses would severly complicate multi-color photometry.  Since the
true SED of an object is unknown, the broad-band photometry itself
provides the best model for the wavelength dependence of the SED,
which is needed to determine the DCR correction.  In practice, this
does not involve making a full correction based on a theoretical
model, but rather by including a color-dependent term in the
astrometric solution \citep[e.g.,][]{pmh+03}.  For SDSS imaging, the
DCR correction is modeled as a linear function of color for stars
bluer than $u-g=3$ in the $u$-band and $g-r=1.5$ in the $g$-band;
redder stars are modeled with constant color.  For the $r$, $i$, and
$z$ bands, the DCR is taken to be a linear function of the $r-i$ color
for all stars.  For more details, see \S~5.3 in
\citet{pmh+03}\footnote{Note that the color dependence was misstated
  in \citet{pmh+03} and is corrected at
  http://www.sdss.org/dr7/products/general/astrometry.html.}.  This
color-dependent modeling of the DCR means that, for purely power-law
sources, the positional agreement between the five
astrometrically-calibrated SDSS bandpasses should not show any
evidence of DCR.

Any residual astrometric offsets will be due to the difference between
the color-dependent correction that is applied to the SDSS astrometry
and the actual DCR for any particular object, which depends on the
distribution of flux {\em within} a bandpass rather than the
distribution of flux {\em across} bandpasses (i.e., the color).  As
such, for our purposes, we can ignore any altitude, temperature,
pressure, or water vapor corrections to the index of refraction of air
since we are only concerned with the {\em relative} DCR with respect
to a presumed model.

\subsection{Predicting SDSS Quasar Positions}

As a result of the broad-band color-dependent DCR correction, we
cannot simply use the effective wavelengths from the composite quasar
spectrum to determine the expected positional offsets of quasars.
Rather, we must compare these offsets with those expected from the
astrometric model.  For our analysis, we will consider two models for
DCR.  One is based on synthetic colors determined from the composite
quasar spectrum \citep[e.g.,][]{rfs+01}, the other is based on the mean
power-law SED of quasars, $\alpha_{\nu}\sim-0.5$.  Accordingly, we
  modify Equation~\ref{eq:R} to
\begin{equation}
R \simeq [R_0({\rm SED})-R_0({\rm color})]\tan(Z)\label{eq:Rnew2}
\end{equation}
or
\begin{equation}
R \simeq [R_0({\rm SED})-R_0({\rm PL})]\tan(Z).\label{eq:Rnew}
\end{equation}

\begin{figure}
\plotone{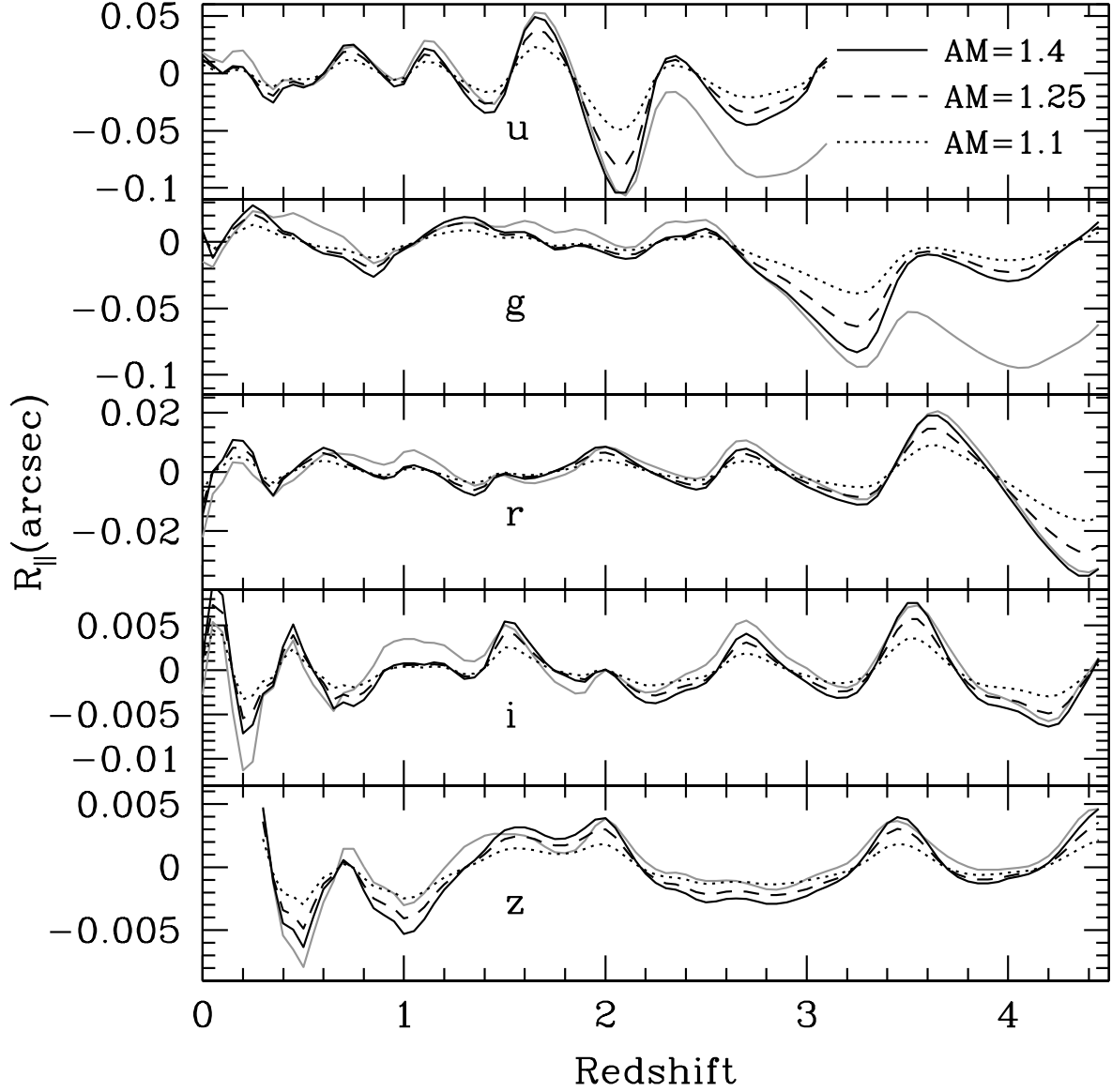}
\caption{Expected change of quasar position as a function of redshift
  and bandpass.  Each panel shows one of the $ugriz$ filters and its
  expected positional change with respect to the position expected
  based on synthetic colors calculated using the \citet{vrb+01} SDSS
  composite quasar spectrum.  Three airmass values are shown in black
  (${\rm AM}=1.4$: {\em solid}; ${\rm AM}=1.25$: {\em dashed}; ${\rm
    AM}=1.1$: {\em dotted}).  The solid gray line is for ${\rm AM}=1.4$,
  but uses an $\alpha_{\nu}=-0.5$ power-law continuum instead of the
  synthetic colors to model the expected DCR.
\label{fig:offsetband}}
\end{figure}

Once we have determined $\lambda_{eff}$ (and thus $R_0$) for both the
composite spectrum and the model in each bandpass as a function of
redshift, positional offsets can be calculated for any airmass
according using Equation~\ref{eq:Rnew2}.  Figure~\ref{fig:offsetband}
shows the predicted offsets for the mean quasar as a function of
redshift and airmass in each of the five SDSS bandpasses.  Here we
make no airmass corrections to the SDSS bandpasses despite the fact
that they are given specifically for ${\rm AM}=1.3$.  As demonstrated
by D.\ Schlegel (unpublished), the 25-35 mas rms astrometric errors
between bandpasses of the SDSS imaging survey \citep{pmh+03} are
comparable to or less than the astrometric offsets due to quasar
emission lines redshifting through the SDSS broad-band filters.

Looking ahead to comparisons with the data, it must be realized that
the DCR effect is entirely in the direction along the ``parallactic
angle'' (or perpendicular to the horizon), thus we adopt the notation
$R_{||}$ in Figure~\ref{fig:offsetband} to indicate positional offsets
in the direction where DCR is applicable.  Our sign convention is that
bluer effective wavelengths result in images higher in the sky than
expected, yielding positive astrometric offsets.

\begin{figure}
\plotone{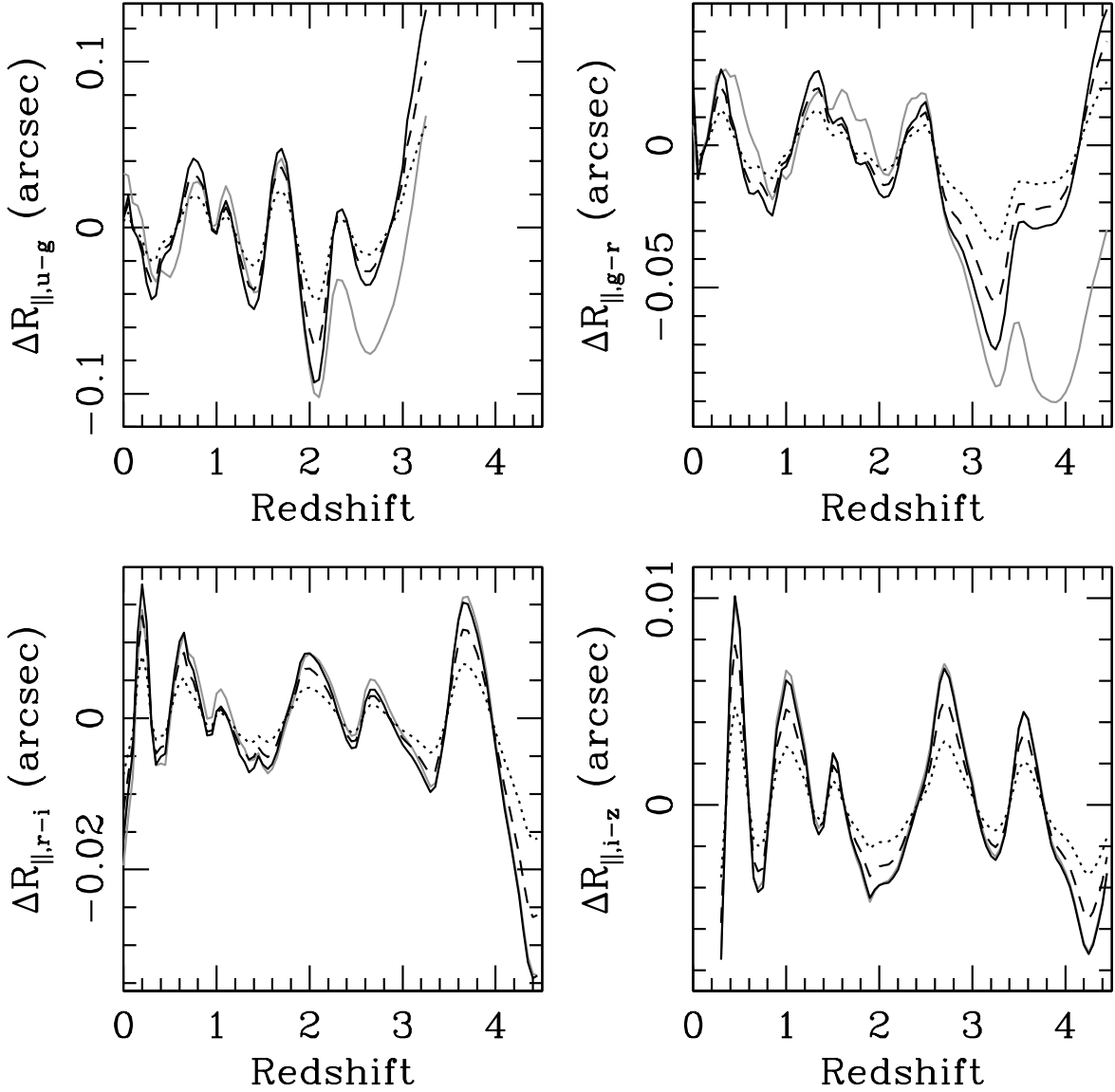}
\caption{Predicted positional differences (in arcseconds) between
  adjacent bandpasses as shown in Fig.~\ref{fig:offsetband}, again as
  a function of redshift and airmass.  The difference in positional
  offsets is represented by $\Delta R_{||,m-n}$ where $m$ and $n$
  represent two bandpasses. For instance, $\Delta R_{||,u-g}$ is the
  expected positional offset of the $u$ band with respect to the $g$
  band.  Line styles and colors are as in Fig.~\ref{fig:offsetband}.
\label{fig:offsetcolor}}
\end{figure}

We will see in \S~\ref{sec:data} that the positions of SDSS objects in
the $ugiz$ bands are given with respect to the $r$ band, which defines
the SDSS astrometric system.  That is the $r$-band positional offsets
do not agree with Figure~\ref{fig:offsetband}, but rather are all
identically zero (by definition).  Thus it will often be more
convenient to examine the relative changes in position between two
bandpasses than the absolute deviation for a single bandpass.  The
difference in positional offsets between two adjacent colors, $\Delta
R_{||,m-n}$ is just $R_{||,m}-R_{||,n}$, where $m$ and $n$ represent
the two bandpasses and.  Figure~\ref{fig:offsetcolor} shows the
resulting $\Delta R_{||,m-n}$ relations as a function of redshift for
adjacent bandpasses at various airmasses.  At sufficiently high
airmass or by averaging multi-epoch observations at low airmass, these
positional offsets can exceed the astrometric errors and can be used
as quasar redshift diagnostics.


\section{Observations}
\label{sec:data}

\subsection{Sloan Digital Sky Survey Data}

While the DCR effect is a well-known phenomenon in astrometry and is
generic to all ground-based data sets, our application of it to
astrometric redshifts is based on data from the Sloan Digital Sky
Survey (SDSS).  The SDSS photometric data is summarized in a series of
papers including \citet{gcr+98}, \citet{yaa+00}, \citet{hfs+01},
\citet{stk+02}, \citet{ils+04}, \citet{gsm+06}, and \citet{tkr+06}.
In \S~\ref{sec:multi}, we will make particular use of the data from
SDSS ``Stripe 82'' \citep{alm+06}, where multiple epochs of SDSS
imaging data have been combined, reducing both the photometric and
astrometric errors.  In addition, we will make use of
spectroscopically-confirmed SDSS quasars that were selected according
to \citet{rfn+02} and tiled according to \citet{blm+03}.  Discussions
of the SDSS filter system and astrometric solutions can be found in
\citet{fig+96}, \citet{slb+02}, \citet{pmh+03}, and also at
http://www.sdss.org/dr7/algorithms/astrometry.html.

\subsection{Single-Epoch Data}

We now compare our theoretical predictions with actual measurements
for the 77,000 quasars from the DR5 quasar catalog \citep{shr+07}.
SDSS astrometry is reported with respect to the $r$ band (corrected to
AM=1).  The positions of the quasars in the other bandpasses were
extracted from the {\tt photoObjAll} in the SDSS
database\footnote{http://cas.sdss.org/dr6/}.  These values are given
in the database as {\tt offsetRA$_x$} (=RA$_x$-RA$_r$cos[Dec$_r$]) and
{\tt offsetDec$_x$} (=Dec$_x$-Dec$_r$), where $x$ represents one of
the $ugriz$ bandpasses.  By definition {\tt offsetRA$_r$} and {\tt
  offsetDec$_r$} are identically zero as the $r$ band (corrected for
atmospheric refraction) defines the SDSS astrometric system.



The DR5 quasar catalog spans a redshift range of 0.08$\leq$z$\leq$5.4
and an airmass range of 1.002$\leq$AM$\leq$1.795.  As we are
interested only in the positional offsets due to DCR, we must first
determine the parallactic angle for each of the observations and
project the observed offsets onto it.  The parallactic angle, $p$, is
given by
\begin{equation}
\sin(p) = \cos(\phi)\frac{\sin {\rm HA}}{\sin(Z)},
\end{equation}
where $\phi$ is the latitude of the observatory and ${\rm HA}$ is the
``hour angle'' of the observation \citep[e.g.,][]{fil82}.  That
portion of the positional offsets parallel to the parallactic angle
comes from a combination of astrometric errors and atmospheric
dispersion, while the portion perpendicular to the parallactic angle
is due only to astrometric errors.

Figure~\ref{fig:offsetplot2} shows the resulting $\Delta R_{||,m-n}$
for the SDSS $ugr$ bandpasses.  While the deviation of the astrometric
offsets from the model is more noticeable at higher airmass and most
SDSS observations are performed at the minimum possible airmass, the
effect is still quite apparent in the ensemble average.


\subsection{Multi-Epoch Data}
\label{sec:multi}

\begin{figure}
\plotone{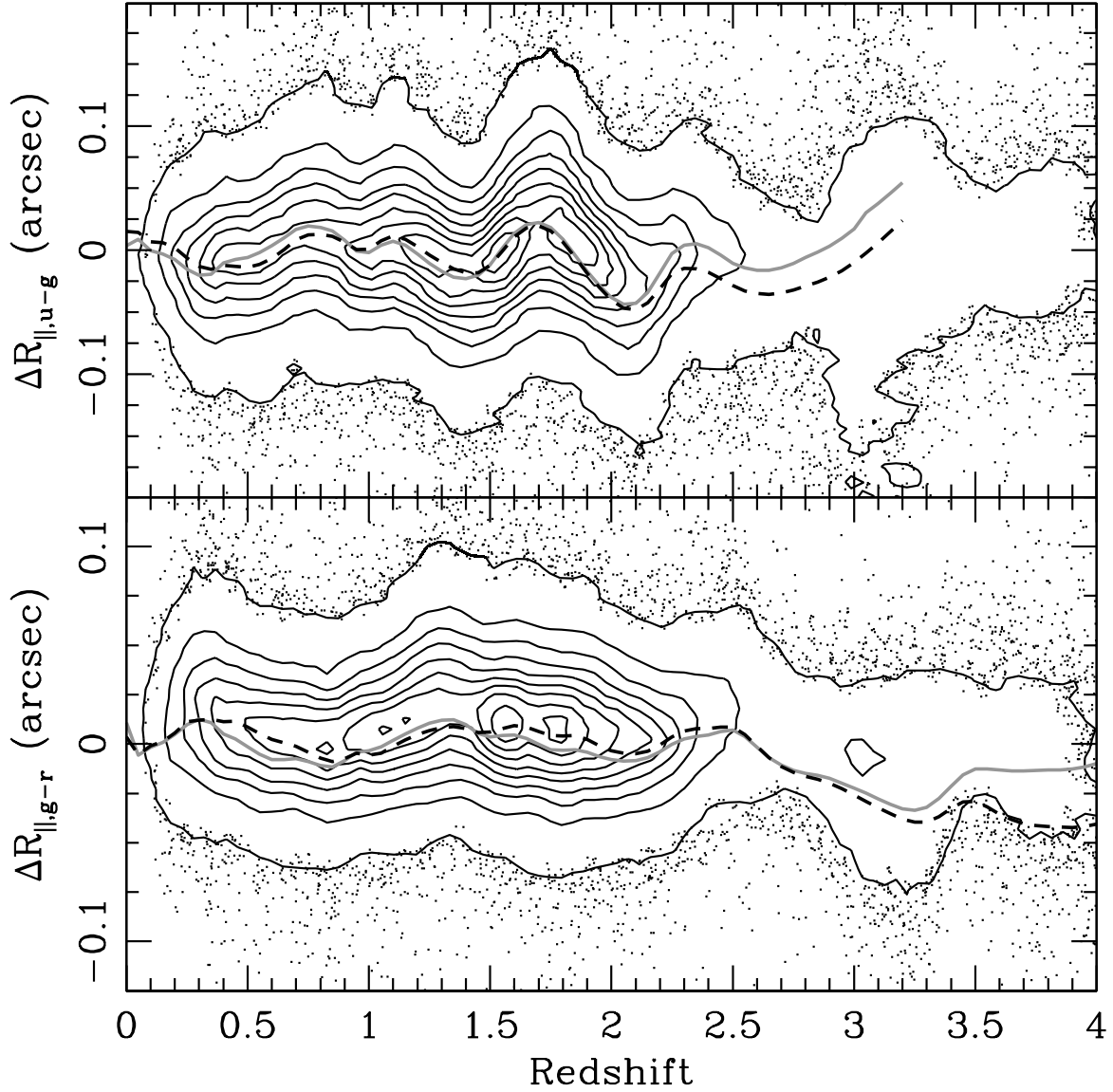}
\caption{Observed positional offsets from emission line induced
  atmospheric dispersion for single-epoch observations of SDSS
  quasars.  Only $\Delta R_{||,u-g}$ and $\Delta R_{||,g-r}$ are
  shown, since they are expected to show a measurable effect.  The
  dashed line indicates the expected offsets from a quasar template
  assuming observations at AM=1.1 and a color-based DCR correction;
  the solid gray line assumes a power-law model for the DCR
  correction.
\label{fig:offsetplot2}}
\end{figure}

Although Figure~\ref{fig:offsetplot2} shows that emission-line-induced
astrometric offsets for quasars are evident in single-epoch SDSS data,
they are not significantly larger than the mean astrometric errors for
individual observations, reducing the utility of the astrometric
offsets for redshift determination.  However, for the SDSS southern
equatorial region (aka, ``stripe 82'') and for the upcoming
multi-epoch synoptic surveys (e.g., Pan-STARRS and LSST), it is
possible to combine data from different epochs in order to reduce
random astrometric errors and increase the signal-to-noise of the
astrometric offsets along the parallactic angle.

To illustrate this improvement we have queried the SDSS stripe 82
database for multiple observations of objects.  While stripe 82 has
dozens of repeat scans, not all of these observations were observed at
airmass high enough to show the DCR effect and not all were observed
at the same airmass.  Thus, we only used observations objects with
${\rm AM}\ge1.2$ and we have combined multiple such observations
together, scaling by $\tan(Z)$ to a mean airmass of 1.3.  This
approach allows us to show not only the benefits of combining the few
dozen epochs available from SDSS stripe 82 data, but also the hundreds
of epochs that will be available in future imaging surveys.

\begin{figure}
\plotone{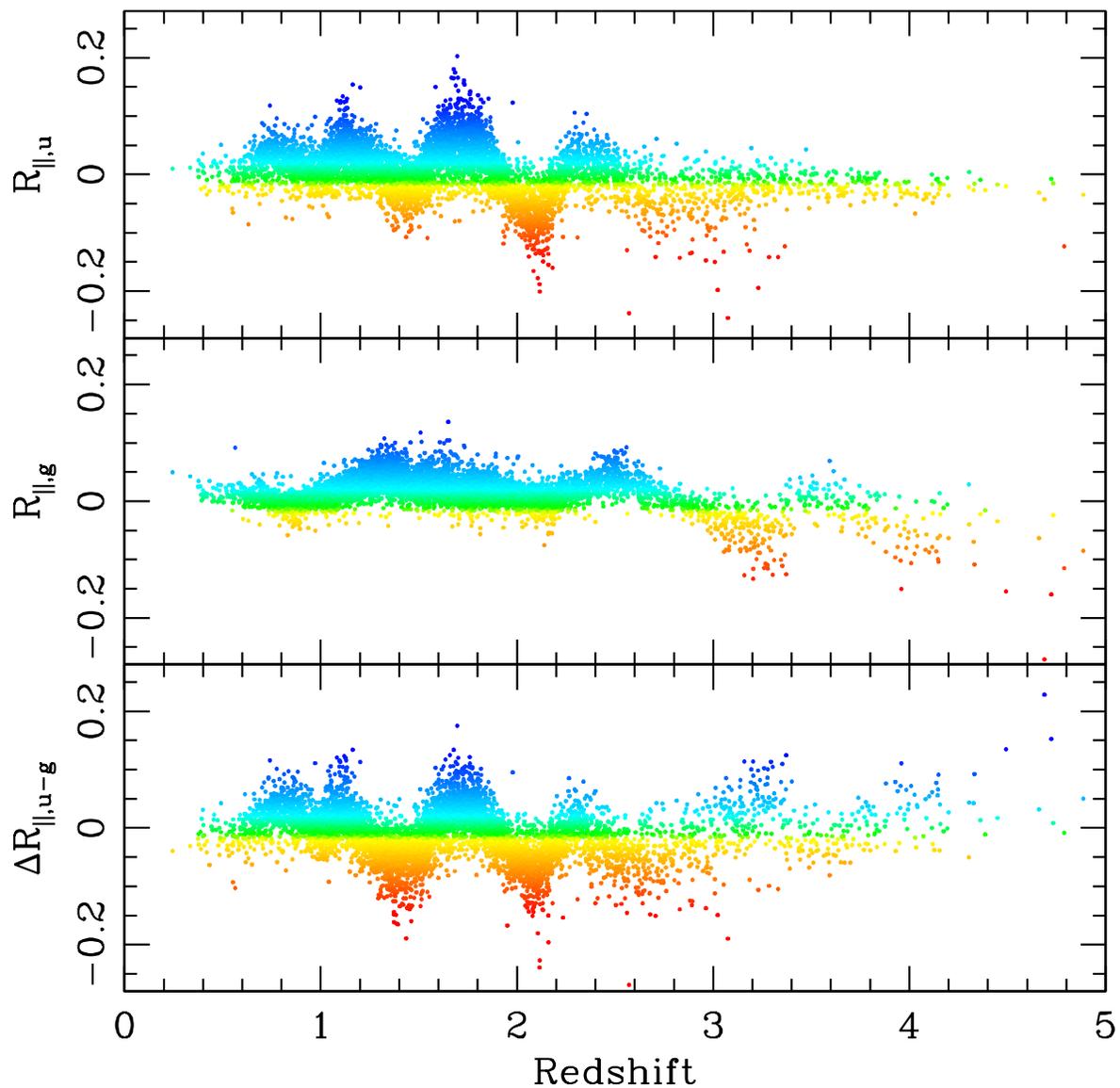}
\caption{Astrometric offsets (in arcseconds) for combined, multi-epoch
  observations of 6430 SDSS quasars in stripe 82.  Points are
  color-coded by the amount of their positional offsets along the
  parallactic angle.  Bluer effective wavelengths lead to more
  positive offsets and are given by bluer points.  The top, middle,
  and bottom panels are for $u$-band offsets, $g$-band offset, and the
  difference of the two, respectively.
\label{fig:qozr}}
\end{figure}

Figure~\ref{fig:qozr} shows the resulting, more accurate, positional
offsets in the $u$, and $g$ bandpasses, along with their difference
for 6430 quasars in SDSS stripe 82.  Comparison of the bottom panel of
Figure~\ref{fig:qozr} and the top panel of
Figure~\ref{fig:offsetplot2} demonstrates that combining multiple
observations has significantly reduced the positional scatter.  The
range of astrometric offsets is nearly 300 mas (e.g., $z\sim1.7$
vs.\ $z\sim 2.1$), as compared with the 25--35 mas astrometric errors
for the single-epoch SDSS data.  While few objects have offsets
considerably larger than 30 mas, over 60\% of the quasars have
astrometric offsets at least this large.  Moreover, by reducing random
scatter with combined multi-epoch data, astrometric offsets smaller
than 30 mas carry useful information.  For example a measured value of
$\Delta R_{||,u-g}=0$ may not uniquely identify the redshift, but it
does suggest redshift ranges that are higly unlikely.

\begin{figure}
\plotone{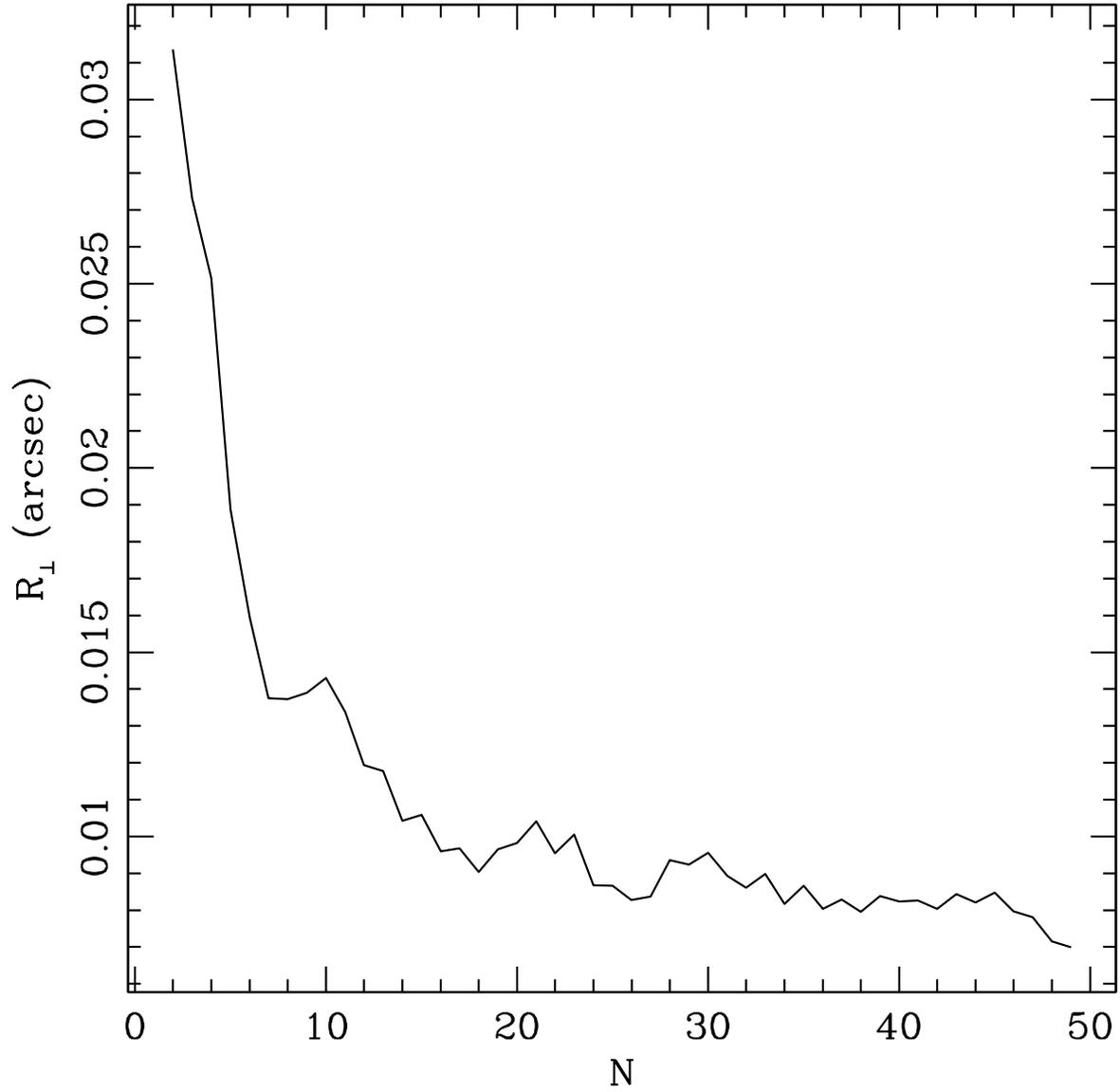}
\caption{Reduction of random astrometric error as a function of number
  of combined epochs.  Single epoch SDSS observations have 25-30 mas
  astrometric errors.  Combining multiple observations can reduce this
  to $\sim$7--8 mas.  These results were derived from bootstrap
  resampling the deviations perpendicular to the parallactic angle for
  50 white dwarf observations.  \label{fig:nvsdx}} \end{figure}

Figure~\ref{fig:nvsdx} illustrates the reduction of random astrometric
errors from 25--35 mas in single epoch observations to under 10 mas
after a few dozen combined observations.  To determine the astrometric
improvement we took the $u$ and $g$ band offsets perpendicular to the
parallactic angle for 50 hot white dwarf observations.  That is we
looked at the astrometric offsets in the direction where we expect
only random errors and not systematic DCR effects.  Those observations
were bootstrap resampled leaving out N objects at a time from 1 to 50
in order to determine the mean improvement as a function of the number
of epochs combined.  While individual observations are noisy, the mean
offets in the direction parallel to the parallactic angle (due to both
signal and noise) are much larger than the perpendicular scatter (due
to noise only).  These differences bode well for the inclusion of
astrometric offsets in photometric redshift estimation and, perhaps,
in object classification (e.g., see the object classification work by
\citealt{lhj+08} based on proper motions of objects in stripe 82).

\section{Astrometric Redshifts}

\begin{figure}
\plotone{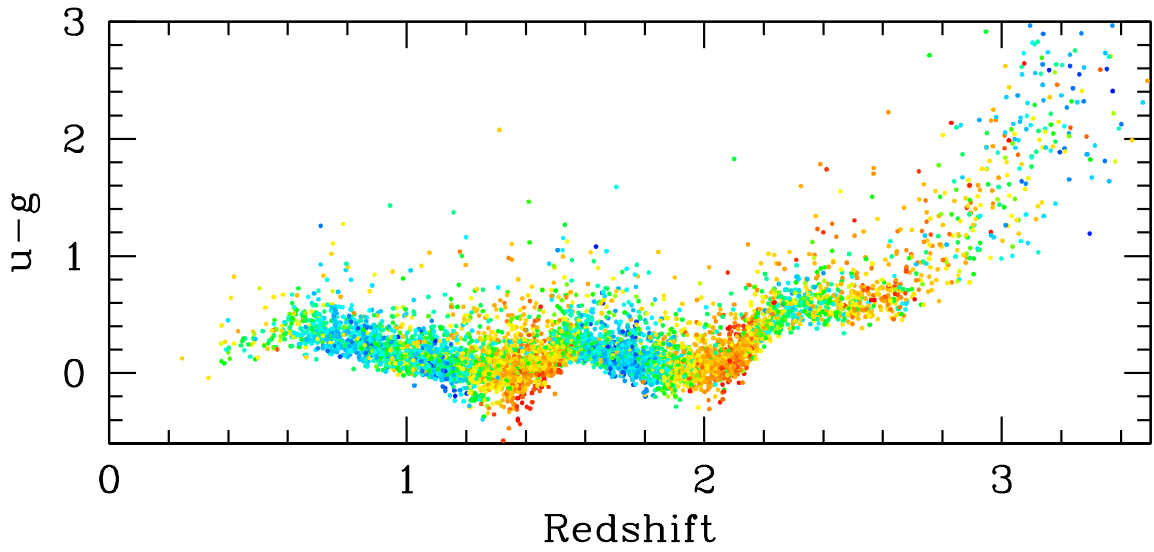}
\caption{$u-g$ color as a function of redshift for SDSS quasars,
  color-coded by the positional offsets from Fig.~\ref{fig:qozr}.
  While an emission line moving through a bandpass produces a color
  change that is roughly symmetric with redshift, the positional
  offsets break this symmetry.
\label{fig:qczr}}
\end{figure}

Since the random astrometric errors for the combined multi-epoch data
are less than the range of astrometric offsets for quasars, these
values can aid in redshift estimation.  Figure~\ref{fig:qczr}
illustrates how DCR can be helpful in this endeavor.  For example,
emission features cause a blueward dip in the $u-g$ color for quasars
with $1.7<z<2.1$.  The $u-g$ color is roughly degenerate for a quasar
at $z=1.7$ and $z=2.1$, which is a source of catastrophic error for
normal photo-$z$'s.  The color-coding of Figure~\ref{fig:qczr} shows
that including the $u-g$ astrometric offset breaks this degeracy by
identifying the color of the $z=1.7$ quasars as being due to the
emission feature on the blue side of the bandpass rather than on the
red side of the bandpass.

Because the astrometric offsets are roughly Gaussian distributed (as
are quasar colors), it is straightforward to include such measurements
in our existing algorithm for redshift estimation as discussed in
\citet{wrs+04}.  We can simply treat the observed $R_{||,u}$ and
$R_{||,g}$ measurements as additional ``colors''.  This method
compares each observed color to the mean (and standard deviation)
color at every redshift, choosing the redshift that minimizes the
$\chi^2$.  

For the 6430 quasars from stripe 82 we have been able to combine
multiple observations, reducing both the photometric and astrometric
errors.  To determine the photo-$z$ improvement due to our
astro-$z$'s, we must first determine the photo-$z$ distribution using
the improved photometry of these objects since we want to distinguish
between astrometric and photometric improvements.  Photo-$z$'s for
these 6430 quasars are first computed using only the five SDSS
magnitudes (and errors).  We perform 5-fold cross-validation whereby
80\% of the sample is used for training and 20\% is used for testing.
Five iterations are used so that photo-$z$'s are estimated for the
full sample.  These 4-color photo-$z$'s provide a baseline for
comparison with redshift estimates that also include astrometric
offsets.

\begin{figure}
\plotone{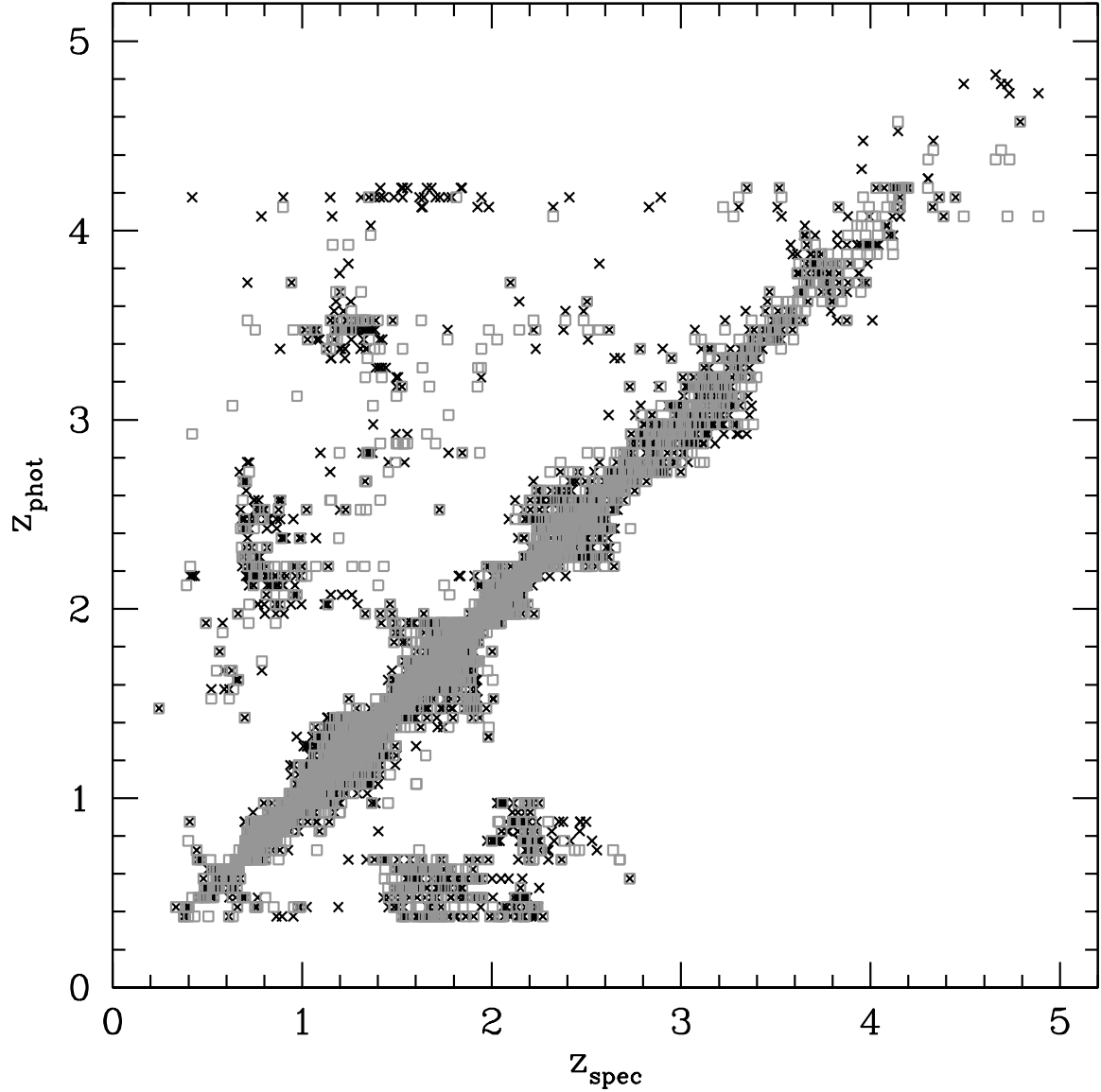}
\caption{Comparison of spectroscopic and photometric redshifts.  Black
  crosses show photo-$z$ only, whereas grey squares are estimates from
  photo-$z$+astro-$z$.  The latter have fewer catastrophic errors and a
  tighter distribution along the diagonal.
\label{fig:zz}}
\end{figure}

\begin{figure}
\plotone{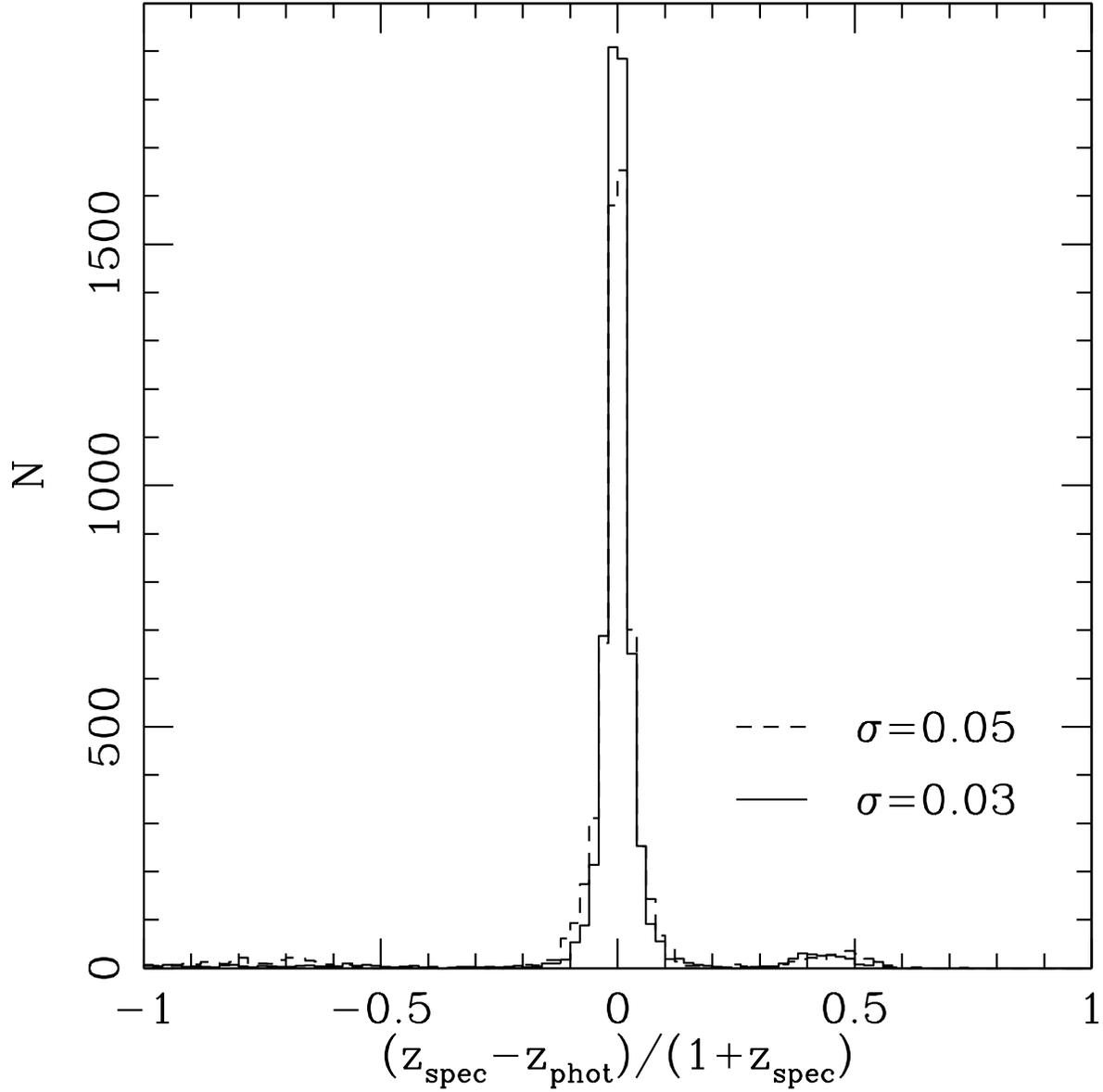}
\caption{Fractional improvement in redshift estimates from the
  inclusion of astrometric offsets (solid) as compared to photometric
  redshifts from colors alone (dashed).  Astro-$z$ information enables
  a significant improvement in accuracy.
\label{fig:dzhist}}
\end{figure}

We then repeat this process, now adding the $R_{||,u}$ and $R_{||,g}$
measurements, treating them as the 5th and 6th colors.  Any
improvement over the 4-color photo-$z$'s must be due to the added
astrometric information.  Figure~\ref{fig:zz} compares the spectropic
versus photometric (and photometric+astrometric) redshifts for the
full sample.  Objects appearing along the diagonal have accruately
estimated redshifts, whereas off-diagonal objects indicate
catastrophic photometric redshift errors \citep{rfs+01}.  While there
are still errors in the photometric+astrometric sample, it can be seen
that inclusion of astrometric information results in somewhat fewer
catastropic errors and a tighter distribution along the diagonal
(i.e., the precision of those objects with accurate photometric
redshifts has been improved).  The improvement can be better seen when
plotting the fractional redshift error as shown in
Figure~\ref{fig:dzhist}.  Adding astrometric redshifts reduces the
fractional error from 5\% to 3\% (excluding catastrophic errors).  It
also increases the fraction of objects within $\Delta z\pm0.3$ by 3\%
and the fraction within $\Delta z\pm0.1$ by 9\%.

We emphasize that this improvement comes simply from treating the
residual astrometric offsets as colors using our existing photo-$z$
algorithm.  In reality, it would make more sense to use the offsets in
a different manner.  In particular, in addition to reducing the
scatter in quasar redshift estimates, Figure~\ref{fig:qczr} suggests
that astro-$z$'s can help break color degeneracies and will help
improve redshift estimates for objects with catastropic errors.  For
example, broad absorption line quasars \citep[e.g.,][]{thr+06} and
dust reddened quasars \citep{rhv+03,mhw+08} can ruin color-based
photometric redshifts algorithms.  High-redshift quasars become very
red in $u-g$ as a result of Lyman-$\alpha$ forest absorption, but
intrinsically blue quasars that are reddened by dust can appear just
as red, which results in catastrophic photo-$z$ errors.  However, the
atmospheric dispersion effect only depends on the distribution of flux
{\em within} the bandpass.  As such, a moderate amount of dust has
little effect on the positional offsets due to strong emission lines.
This suggests that astro-$z$'s may perform better when used as priors
in advance of a standard photo-$z$ algorithm.

\section{Future Work}

\subsection{Other Emission-Line Objects}

\begin{figure}
\plotone{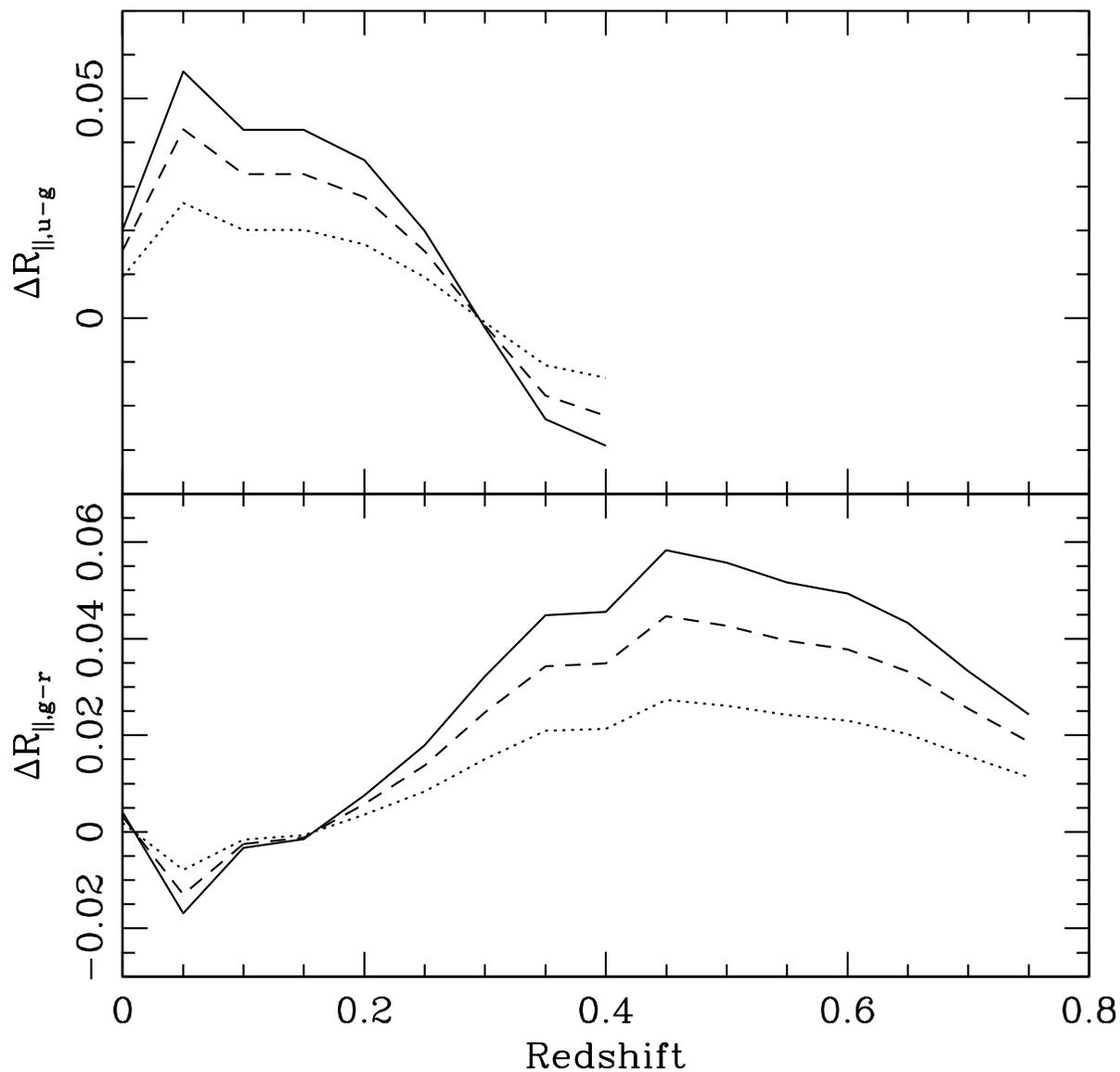}
\caption{Positional offsets (in arcseconds) for type 2 quasars, based on a composite
  type 2 spectrum from \citet{zsk+03}.  Results for three airmasses
  are shown, labeled as in Fig.~\ref{fig:offsetband}.  While the
  offsets for type 2 quasars are smaller than for type 1 quasars, the
  falling trend in $R_{||,u-g}$ coupled with the rising trend in
  $R_{||,g-r}$ is encouraging with regard to type 2 redshift
  prediction.
\label{fig:offsetcolortype2}}
\end{figure}

We have concentrated on the applications of astrometric offsets
specifically for quasars; however, the method is potentially
applicable to other emission-line objects (e.g., type 2 quasars and
supernovae).  For type 2 quasars the emission lines are narrow, but
their strength means that we can expect to see measurable astrometric
offsets that are a function of redshift.  Using a template type 2
quasar spectrum \citep[e.g.,][]{zsk+03}, we can estimate the
astrometric offsets due to strong emission lines in type 2 quasar
spectra.  Figure~\ref{fig:offsetcolortype2} shows that, over the range
$0.05<z<0.4$, $R_{||,u-g}$ falls by almost 90 mas, while $R_{||,u-g}$
rises by 75 mas.  Thus redshifts for type 2 quasars with $z\le0.4$
could be distinguished at roughly the same level of accuracy as type 1
quasars.

It may also be possible to estimate redshifts for supernovae using
astrometric offsets.  While their SEDs are highly time-dependent and it
would not be possible to combine multiple epochs, if the distinctive
emission features at each epoch produced redshifts estimates that were
in agreement, these astro-$z$'s might be as reliable as they would be if
the SEDs were not time-dependent.  Most SNe spectral templates don't
cover enough of the UV to properly estimate their positional offsets
with redshift, but based on the existing coverage longward of 3000\AA,
it would appear that the range of $u$- and $g$-band offsets is as
large as $0\farcs15$---certainly enough to make an initial guess at
the redshift.

\subsection{Object Classification}

Just as the nature of the atmospheric offsets allows improvements in
redshift estimation, the offsets could also be utilized in object
classification itself.  While astrometric solutions attempt to remove
DCR, they can only do so at the level of broad-band photometric
accuracy, whereas DCR is sensitive to features smaller than the width
of the photometric bandpasses.  As such, objects with distinctive SEDs
may stand out enough in astrometric offset space to aid in object
classification.  While astrometric offsets alone are not likely to be
useful for object classification, coupling these astrometric
measurements with color and variability information within the context
of modern Bayesian classification algorithms \citep{rmg+08} could
prove very fruitful in terms of the accuracy of object classification
from broad-band imaging data.

\subsection{Upcoming Imaging-only Surveys}

While most surveys aim to obtain observations as close to the meridian
as possible (and thus at the lowest airmass possible), any object with
declination different from the latitude of the observatory will
necessarily be observed at ${\rm AM}>1$.  Thus any imaging survey will
naturally yield observations that can be used with our approach.  As
refraction increases with airmass, 
one should ask whether future surveys may wish to obtain
a fraction of their observations at relatively high airmass.  Such
observations will result in a loss of photons (particularly in the
ultraviolet), so one must consider the details of the tradeoff between
low and high airmass observations.

\begin{figure}
\plotone{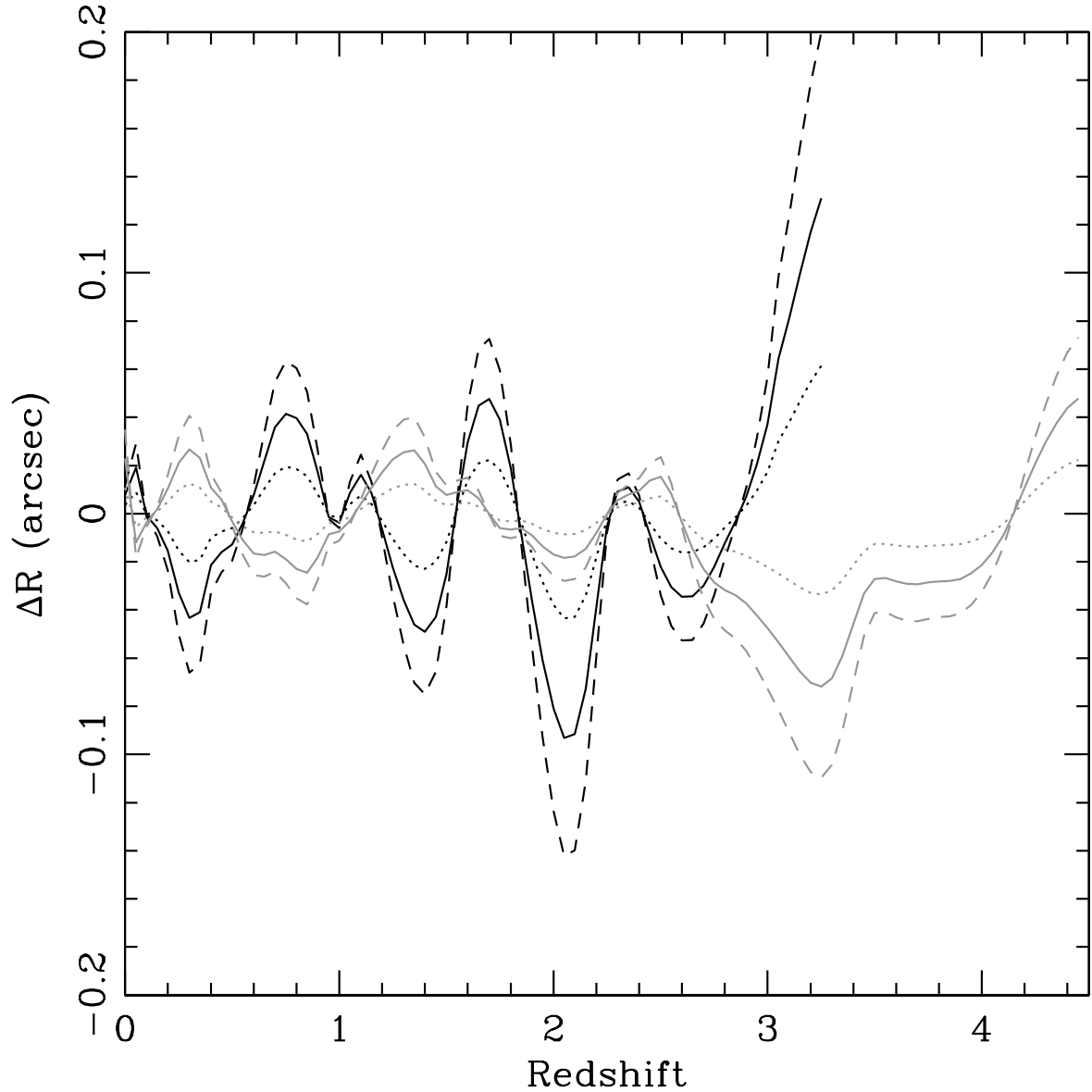}
\caption{Positional offsets as a function of quasar redshift. $\Delta
  R_{||,u-g}$ is shown in black, $\Delta R_{||,g-r}$ in gray.  Solid
  lines give ${\rm AM}=1.4$, dashed lines are ${\rm AM}=1.8$, and
  dotted lines are ${\rm AM}=1.1$.  It is clear that ${\rm AM}=1.1$
  does not provide significant leverage in resolving quasar redshift
  degeneracies, but nor is there significant gain in going from ${\rm
    AM}=1.4$ to ${\rm AM}=1.8$, suggesting moderate airmass
  observations.
\label{fig:amtest}}
\end{figure}

We have already shown in Figure~\ref{fig:nvsdx} that combining just a
few observations can go a long way towards reducing random astrometric
errors below the level of anomalous DCR offsets for quasars.  We next
ask whether there is utility in making observations at very high
airmass in order to increase the snr of the anomalous DCR effect.  In
Figure~\ref{fig:amtest} we overplot the expected offsets for quasars
in both the $u$- and $g$-bands for airmasses 1.1, 1.4, and 1.8.  It
can be seen that an airmass of 1.1 is not sufficient for breaking
redshift degeneracies, which implies that some level of concerted
effort will be needed to obtain observations off of the meridian.  At
an airmass of 1.4, on the other hand, there is sufficient range to
clearly indentify certain redshifts, even considering the scatter of
the measurements as compared to these theoretical distributions.
Extending to ${\rm AM}=1.8$, however, seems to offer little
improvement, perhaps with the exception of the $z\sim2.1$ and
$z\sim3.3$.

A full cost-benefit analysis of the trade-offs between more
observations and higher airmass is beyond the scope of this paper as
such analysis is highly system dependent, but it would seem that of
order 10 observations at a moderate airmass of $\sim1.4$ would be
sufficient to yield significant improvement in photometric redshifts
for quasars without causing significant problems for observational
cadence or data reduction.

While the gains in photometric redshift accuracy due to the addition
of astrometric information shown herein are modest ($\sim$9\%), they
undersell the utility of astrometric redshifts for the next generation
of imaging surveys which will be able to combine hundreds instead of
tens of epochs of data, each epoch with astrometric accuracy superior
to that from the SDSS.  For example, at $r=22$ the LSST astrometric
errors are expected to be $\sim$15 mas and there will exist over
200 epochs of data \citep{lsst08}.

\section{Conclusions}

Astrometrists have long been dealing with the dispersive nature of our
atmosphere when performing broad-band photometry and astrometry.  Here
we show that the dispersive nature of our atmosphere can actually be a
boon rather than a burden.  Objects with strong emission lines produce
measurable astrometric offsets (with respect to an expected model) in
the ensemble average even when observed at relatively low airmass.
The signal-to-noise of the effect can be maximized by either combining
many observations at low airmass or fewer observations at higher
airmass.  With well-measured astrometric offsets, it is possible to
both break degeneracies in photometric redshift estimates of quasars
and to reduce their overall scatter.  For combined, multi-epoch data
from the SDSS stripe 82 region, adding astrometric information
increases the fraction of quasars with correct photometric redshifts
by as much as 9\% ($\Delta z \pm0.1$).  Given the lack of
spectroscopic capabilities for future imaging surveys, the community
would do well to build astrometric redshifts into their survey
planning.

\acknowledgments

GTR acknowledges support from an Alfred P. Sloan Research Fellowship.
We thank \v{Z}eljko Ivezi\'{c}, Jeff Munn, and Don Schneider for
comments on the manuscript and Bob Hindsley for early work on DCR in
the SDSS astrometric system.  Funding for the SDSS and SDSS-II has
been provided by the Alfred P. Sloan Foundation, the Participating
Institutions, the National Science Foundation, the U.S. Department of
Energy, the National Aeronautics and Space Administration, the
Japanese Monbukagakusho, the Max Planck Society, and the Higher
Education Funding Council for England. The SDSS is managed by the
Astrophysical Research Consortium for the Participating
Institutions. The Participating Institutions are the American Museum
of Natural History, Astrophysical Institute Potsdam, University of
Basel, Cambridge University, Case Western Reserve University,
University of Chicago, Drexel University, Fermilab, the Institute for
Advanced Study, the Japan Participation Group, Johns Hopkins
University, the Joint Institute for Nuclear Astrophysics, the Kavli
Institute for Particle Astrophysics and Cosmology, the Korean
Scientist Group, the Chinese Academy of Sciences (LAMOST), Los Alamos
National Laboratory, the Max-Planck-Institute for Astronomy (MPIA),
the Max-Planck-Institute for Astrophysics (MPA), New Mexico State
University, Ohio State University, University of Pittsburgh,
University of Portsmouth, Princeton University, the United States
Naval Observatory, and the University of Washington.



\clearpage

\end{document}